\documentclass[twoside]{LCWS11}
\usepackage[latin1]{inputenc}
\usepackage[pdftex]{graphicx,epsfig,color}
\usepackage{wrapfig,rotating}
\usepackage{amssymb,amsmath,array}
\usepackage{cite}
\usepackage{units}
\usepackage{heppennames2}
\usepackage{subfig}
\DeclareGraphicsExtensions{.pdf, .png}

\begin{document}
\newcommand{\epem}{\ensuremath{\mathrm{\Pep\Pem}}\xspace}
\newcommand{\abinv}{\ensuremath{\mathrm{ab}^{-1}}\xspace}
\newcommand{\pT}{\ensuremath{p_\mathrm{T}}\xspace}

\newcommand{\gghadrons}{\ensuremath{\upgamma\upgamma \rightarrow \mathrm{hadrons}}\xspace}
\newcommand{\clicsid}{CLIC\_SiD\xspace}
\newcommand{\micron}{\ensuremath{\upmu\mathrm{m}}}
\newcommand{\radlen}{\ensuremath{X_{0}}\xspace}
\newcommand{\radlenfrac}{\ensuremath{X/X_{0}}\xspace}
\newcommand{\nuclen}{\ensuremath{\lambda_{\mathrm{I}}}\xspace}
\newcommand{\degrees}{\ensuremath{^{\circ}}\xspace}
\newcommand{\rmlad}{\ensuremath{_{\mathrm{ladder}}}}
\newcommand{\mm}[1]{\ensuremath{_{\mathrm{#1}}~\mathrm{[mm]}}}
\newcommand{\mic}[1]{\ensuremath{_{\mathrm{#1}}~\mathrm{[\micron]}}}
\newcommand{\mumu}{\ensuremath{\upmu\upmu}\xspace}
\newcommand{\nuenuebar}{\ensuremath{\PGne\PAGne}\xspace} 
\newcommand{\mpmm}{\ensuremath{\PGmp\PGmm}\xspace}  
\newcommand{\nunubar}{\ensuremath{\PGn\PAGn}\xspace}   
\newcommand{\tptm}{\ensuremath{\PGtp\PGtm}\xspace} 
\newcommand{\gamgam}{\ensuremath{\upgamma\upgamma}\xspace}
\newcommand{\ww}{\ensuremath{\PWp\PWm}\xspace} 
\newcommand{\zz}{\ensuremath{\PZz\PZz}\xspace} 
\newcommand{\wwz}{\ensuremath{\PWp\PWm\PZz}\xspace} 
\newcommand{\zzz}{\ensuremath{\PZz\PZz\PZz}\xspace} 
\newcommand{\zhsm}{\ensuremath{\PH\PZz}\xspace}  
\newcommand{\hbb}{\ensuremath{\PH\to b\bar{b}}\xspace}
\newcommand{\hcc}{\ensuremath{\PH\to c\bar{c}}\xspace}
\newcommand{\hmumu}{\ensuremath{\PH\to \mumu}\xspace}
\newcommand{\guineapig}{\textsc{GuineaPig}\xspace}
\newcommand{\mokka}{\textsc{Mokka}\xspace}
\newcommand{\marlin}{\textsc{Marlin}\xspace}
\newcommand{\geant}{\textsc{Geant4}\xspace}
\newcommand{\slic}{\textsc{SLIC}\xspace}
\newcommand{\lcsim}{\texttt{org.lcsim}\xspace}
\newcommand{\pythia}{\textsc{PYTHIA}\xspace}
\newcommand{\whizard}{\textsc{WHIZARD}\xspace}
\newcommand{\pandora}{\textsc{PandoraPFA}\xspace}
\newcommand{\fastjet}{\textsc{FastJet}\xspace}
\newcommand{\tmva}{TMVA\xspace}
\newcommand{\roofit}{RooFit\xspace}
\title{Flavour Tagging at CLIC} 
\author{Tomas Lastovicka$^1$ Katja Seidel$^{2,3}$ Jan Strube$^4$
\vspace{.3cm}\\
1- Institute of Physics, Academy of Sciences of the Czech Republic\\
Prague - Czech Republic
\vspace{.1cm}\\
2- Max-Planck-Institut f\"ur Physik\\
Munich - Germany
\vspace{.1cm}\\
3- Excellence Cluster 'Universe', TU M\"unchen\\
Garching - Germany
\vspace{.1cm}\\
4- CERN\\
Meyrin - Switzerland
}


\maketitle

\begin{abstract}
We present the performance of the LCFI flavour tagging package in a realistic CLIC environment. The application is demonstrated on the examples of the measurement of the cross section times branching ratio of light Higgs decays to b and c quarks at \unit[3]{TeV}, a study of heavy Higgs decays at \unit[3]{TeV} and of top pair production at \unit[500]{GeV}. All studies are based on full detector simulation with a realistic account of the machine-induced background at CLIC.
\end{abstract}

\section{Introduction}
The CLIC accelerator provides unique opportunities for precision physics, due to the large production cross section of Higgs bosons at higher energies, and the increased energy reach compared to the ILC could prove to be an essential advantage in studying Supersymmetry. At the same time, the difficult machine environment bears its own challenges for precision physics. We demonstrate that the approach to flavour tagging that was developed for the ILC provides excellent results also at a CLIC machine at \unit[500]{GeV} and at \unit[3]{TeV}.

The LCFI flavour tagging package~\cite{LCFI} has been developed for the ILC and is based on the ZVTOP algorithm invented at SLD. We will give here only a brief recap and focus on the differences to the ILC. After clustering the reconstructed event into jets, the ZVTOP algorithm finds secondary vertices from the tracks in a given jet. In contrast to LEP and ILC analyses, where the Durham jet clustering algorithm is employed, it is found that the CLIC environment requires the use of $k_t$-style algorithms developed for hadronic environments. The performance of the vertex fitting routines on primary vertices is demonstrated in the CLIC\_SiD detector concept.

Information from the secondary vertices is used together with other jet-based variables as the input to neural networks for the flavour tagging.
Accounts of the performance of the flavour tagging in the measurement of the cross section times branching ratio of light Higgs decays to b and c quarks at \unit[3]{TeV}~\cite{lcd:2011-036}, a study of heavy Higgs decays at \unit[3]{TeV}~\cite{Battaglia:2010in} and of top pair production at \unit[500]{GeV}~\cite{LCD-2011-026} are given in Section~\ref{sec:physicsPerf}. All analyses use a full GEANT4 detector simulation and take into account realistic CLIC backgrounds.
\section{Vertex Occupancies}
\begin{figure}
	\centering
	\subfloat[barrel]{\includegraphics[width=0.43\linewidth]{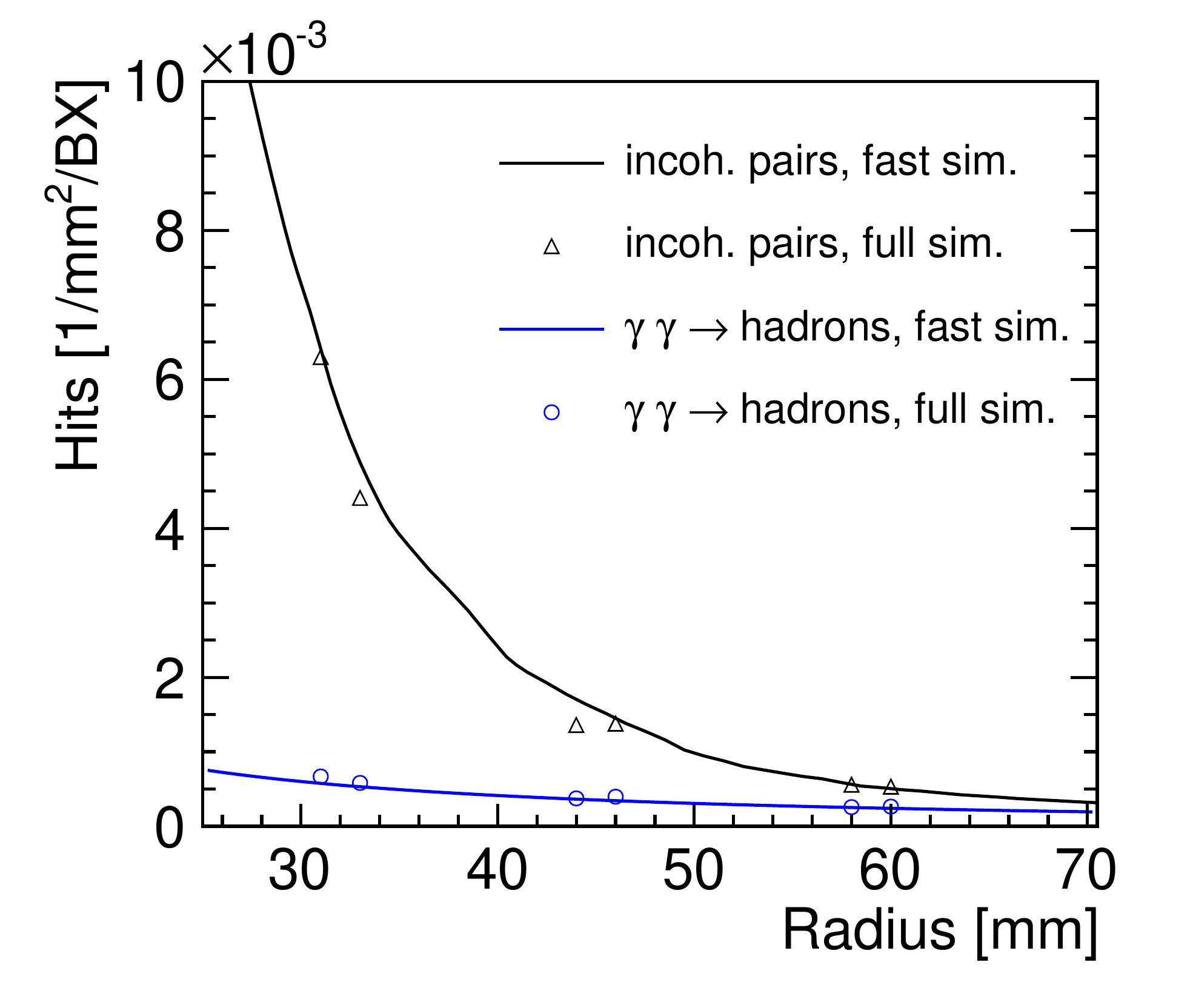}\label{fig:occupancies_barrel}} \hfill
	\subfloat[forward disks]{\includegraphics[width=0.43\linewidth]{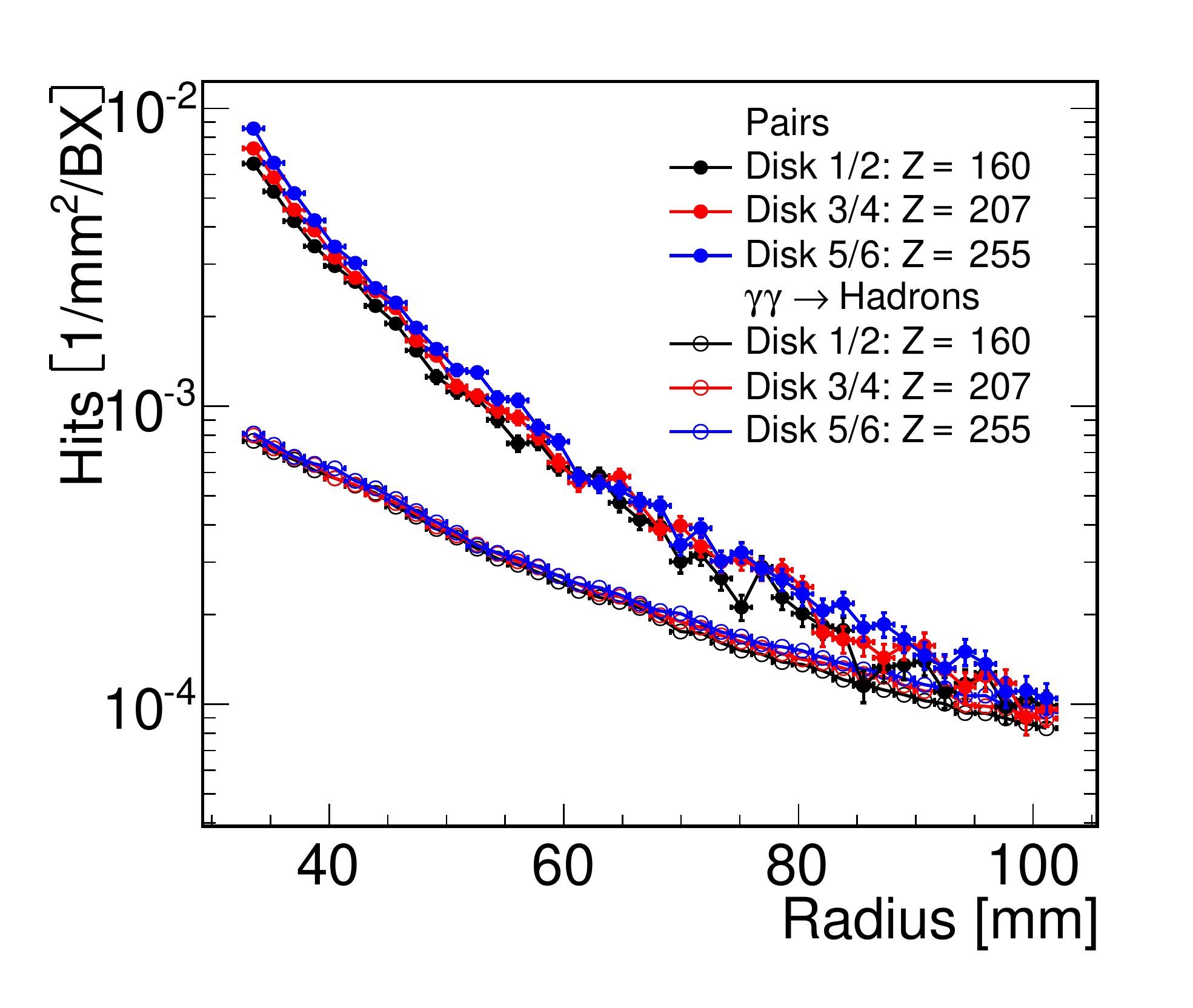}\label{fig:occupancies_ftd}}
	\caption{Average hit densities in the CLIC\_ILD barrel (left) and forward (right) vertex detectors for particles originating from incoherent electron-positron pairs and from \gghadrons. For the barrel region, the full simulation of the detector response is compared to a fast parametric tracking of the primary particles in the magnetic field. For the forward region, the results are shown for the full simulation only. Safety factors for the simulation uncertainties and cluster formation are not included.}
	\label{fig:vtx_occupancies}
\end{figure}
The high fields of the CLIC beams produce several hadronic events in photon collisions per bunch crossing~\cite{lcd:2011-DannheimSailerBgrNote}.
This background from \gghadrons and incoherent electron-positron pairs leads to high occupancies of the inner layers of the vertex detectors, potentially posing challenges to identifying secondary decay vertices close to the interaction point (IP). Figure~\ref{fig:vtx_occupancies} shows the occupancies from these two background sources in the barrel and forward regions of the vertex detector of the CLIC\_ILD concept at \unit[3]{TeV}. The position of the vertex detector layers is indicated by the x-coordinate of the black empty triangles or the blue circles. To reduce this background the detector concepts at CLIC make extensive use of time stamping, assuming \unit[10]{ns} in the silicon detectors. While the occupancies in the inner double layer are several times higher than in the other layers, mainly due to background from incoherent electron-positron pairs, these hits have negligible impact on the track reconstruction efficiencies~\cite{lcd:grefesidtracking2011,LCD:2011-013}.

\section{Jet Clustering}
\begin{figure}
	\centering
	\includegraphics[width=0.43\linewidth]{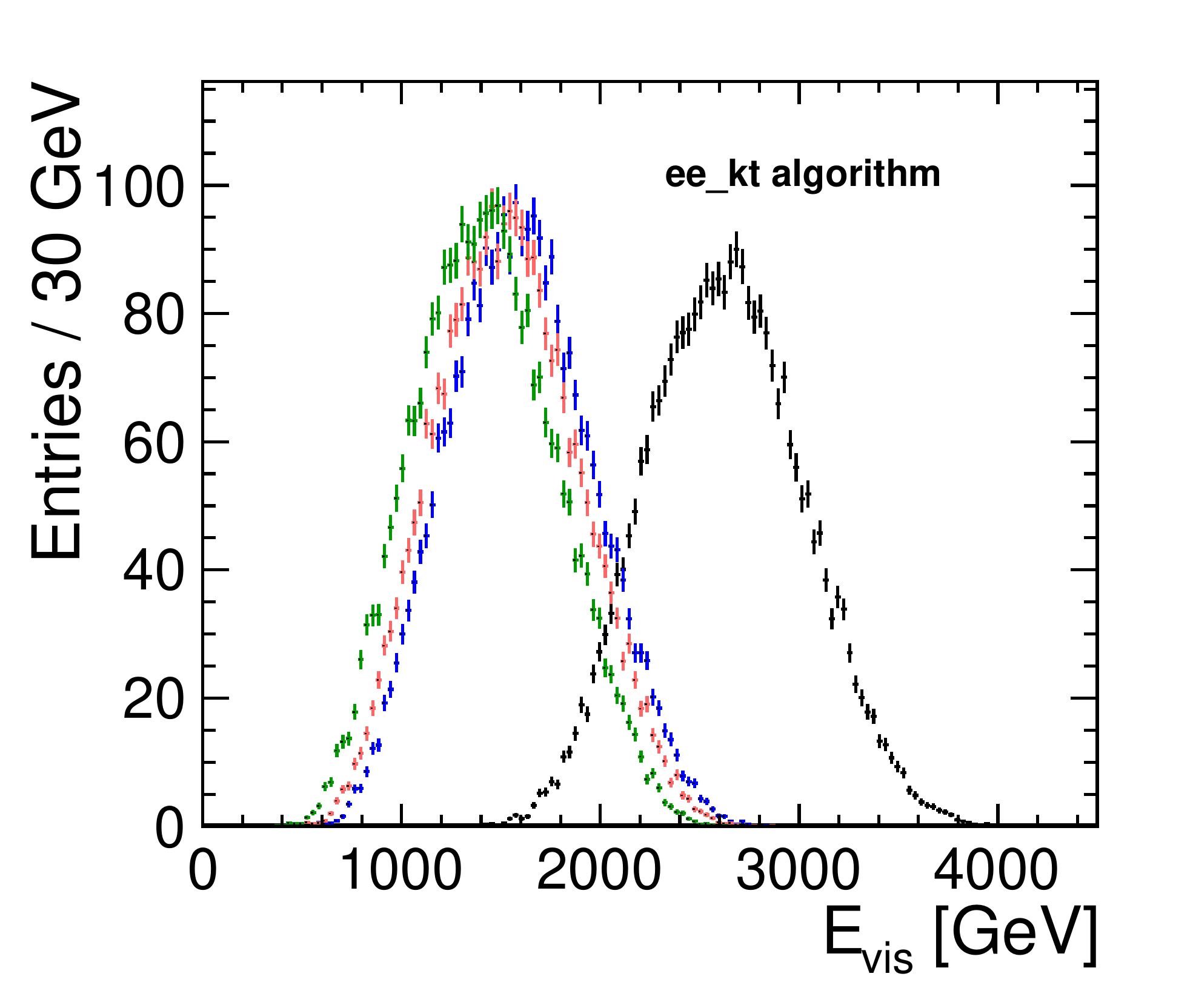}
	\includegraphics[width=0.43\linewidth]{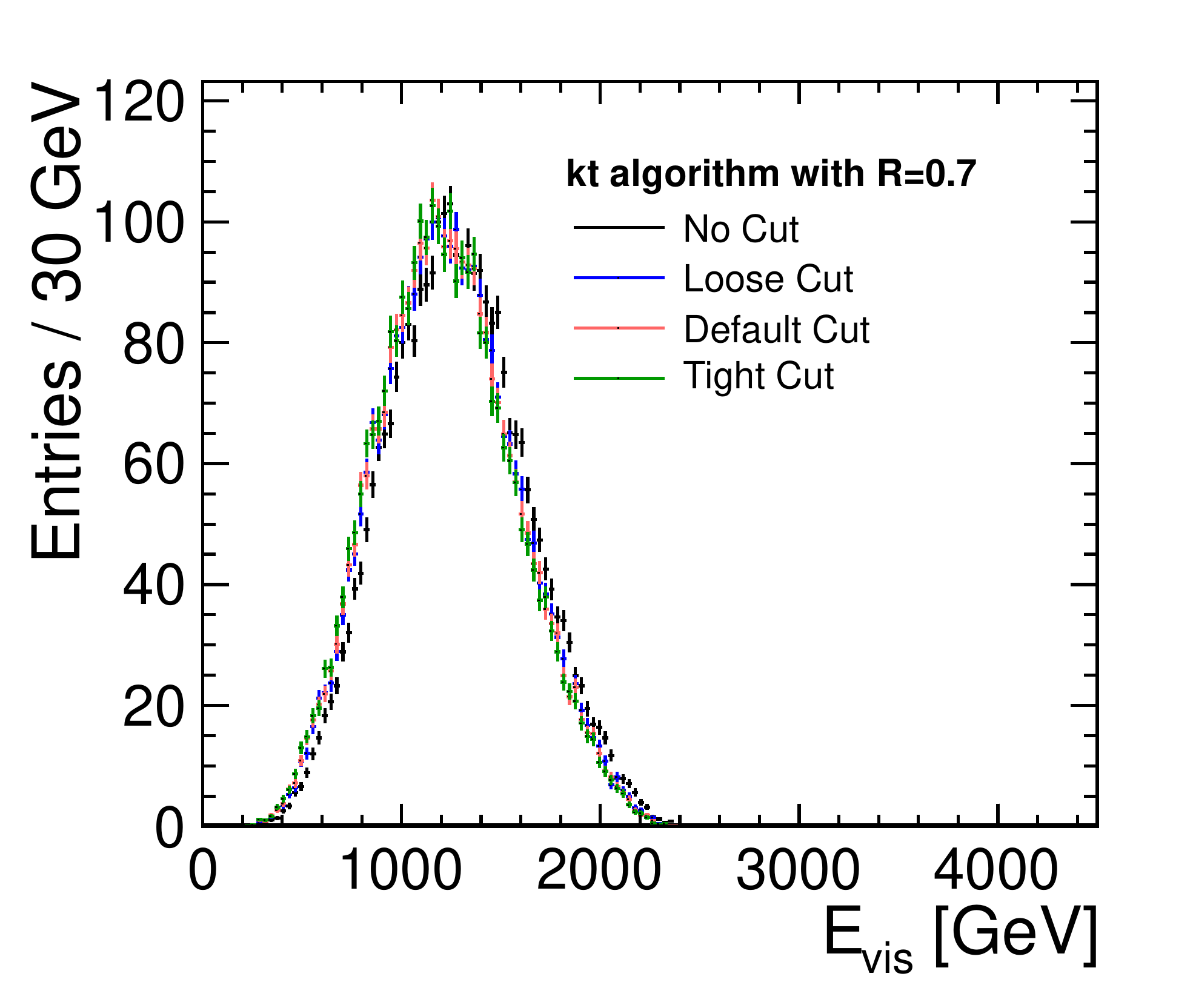}
	\caption{The visible energy in events with two jets and missing energy for different timing cuts (\texttt{tight}, \texttt{default} and \texttt{loose}) and without timing cuts on the reconstructed particles for the Durham (ee\_kt) algorithm (left) and the $k_t$-algorithm (right).}
	\label{fig:jetFinding}
\end{figure}
The LCFI package uses the ZVTOP vertex finding algorithm to identify regions where multiple tracks originate from the decay of a long-lived secondary particle. The seed for such a vertex candidate is found by pairwise overlapping all tracks within a jet. The performance of finding vertices with only one track is greatly enhanced by using the jet axis. However, the beam-induced background from \gghadrons is causing LEP-style clustering algorithms to fail. At a \unit[3]{TeV} CLIC machine, $k_t$ algorithms developed for hadron machines are more appropriate. Figure~\ref{fig:jetFinding} shows the energy that is clustered in jets for a Durham-style algorithm used at LEP (ee\_kt), and for a $k_t$ algorithm as it is applied in the LHC experiments. The different colors show how the jet energy changes with the different timing cuts on the reconstructed particles. The effect of the \gghadrons background is clearly visible. As this background is more dominant in the forward regions of the detector, this clearly has an effect on the found jet axis. Events were clustered into two jets, and while the Durham algorithm clusters all particles into the jets, the $k_t$ algorithm clusters particles in the far forward region into an additional beam-jet that contains most of the background. This reduces a potential bias on the jet axis, which is used in the jet flavour tagging.

\section{Primary Vertex Resolution}
\begin{figure}
	\centering
	\includegraphics[width=.43\linewidth]{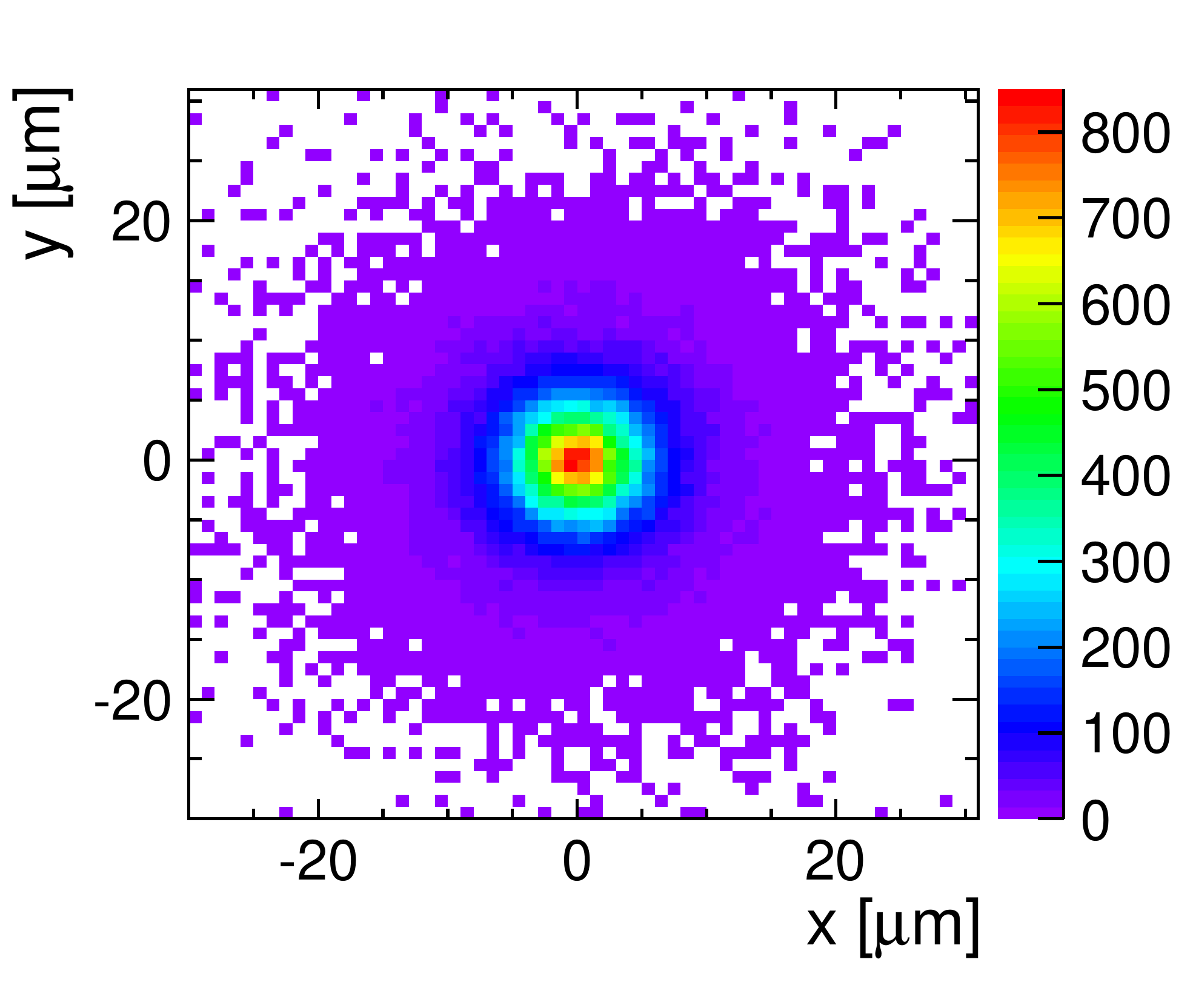}\hfill
	\includegraphics[width=.43\linewidth]{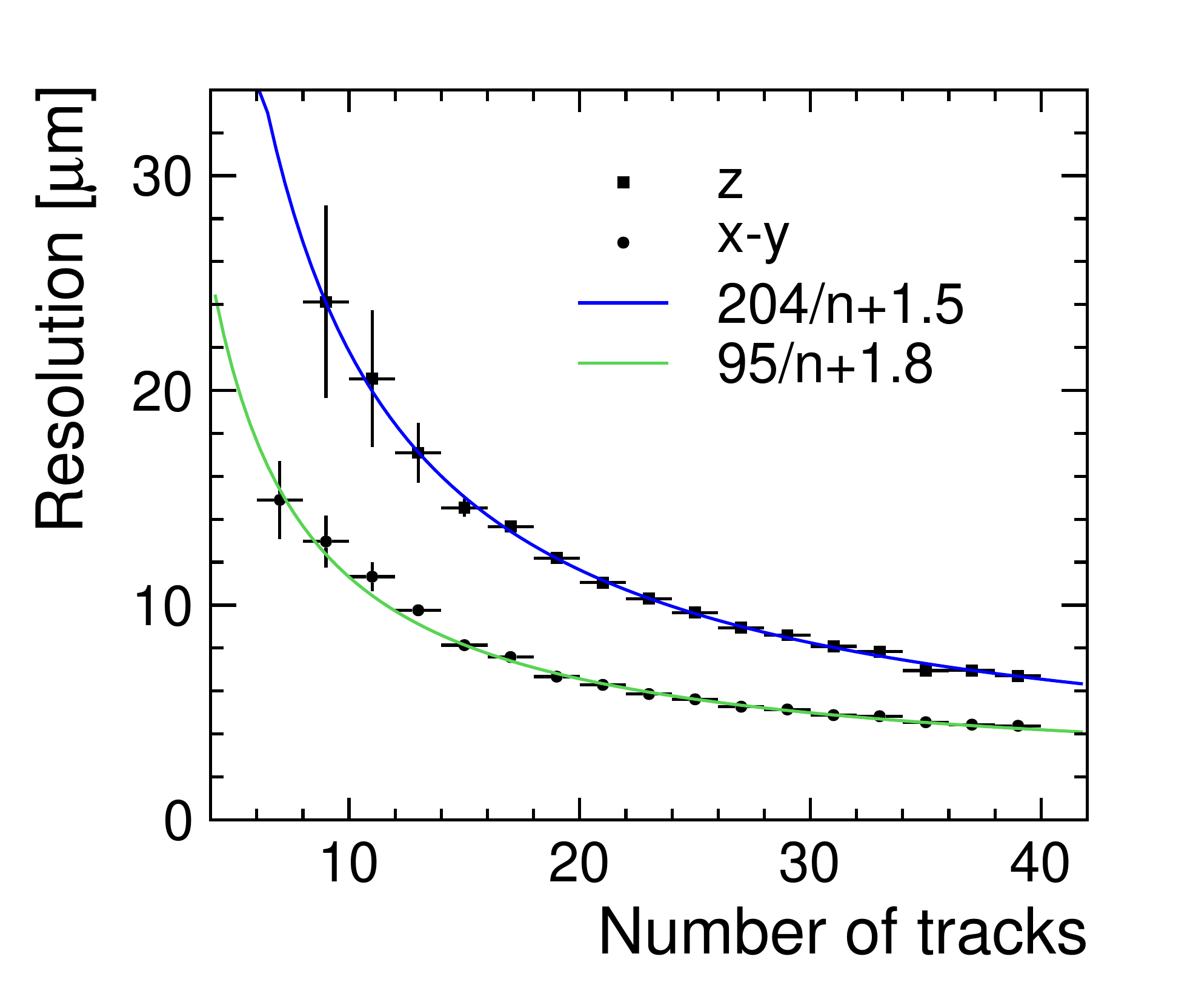}
	\caption{Location in the x-y plane of the primary vertices found in an event (left) and resolution of the primary vertices versus the number of tracks attached to the vertex (right).}
	\label{fig:primary_vertex}
\end{figure}
The secondary vertex decay length is one of the most discriminating variables between b-, c- and light jets. We use the known primary vertex position at CLIC to judge the performance of the vertex reconstruction. Figure~\ref{fig:primary_vertex} (left) shows the position of primary vertices with more than 20 tracks in di-jet events in the x-y plane. The primary vertex in the simulation is always at the origin of the detector coordinate system. Figure~\ref{fig:primary_vertex} (right) shows the vertex resolution of the primary vertex in the x-y plane (solid points) and along the z axis (solid squares) versus the number of tracks in the vertex. The distribution of the reconstructed vertex position was fitted with two Gaussian functions and the shown resolution is the weighted mean of the two fitted functions. The resolution of secondary vertices can be deduced from this figure, as it depends primarily on the number of tracks.

\section{Flavour Tagging Performance on Physics Samples}
\label{sec:physicsPerf}
\subsection{Performance at \unit[3]{TeV}}
\begin{figure}
	\centering
	\includegraphics[width=.43\linewidth]{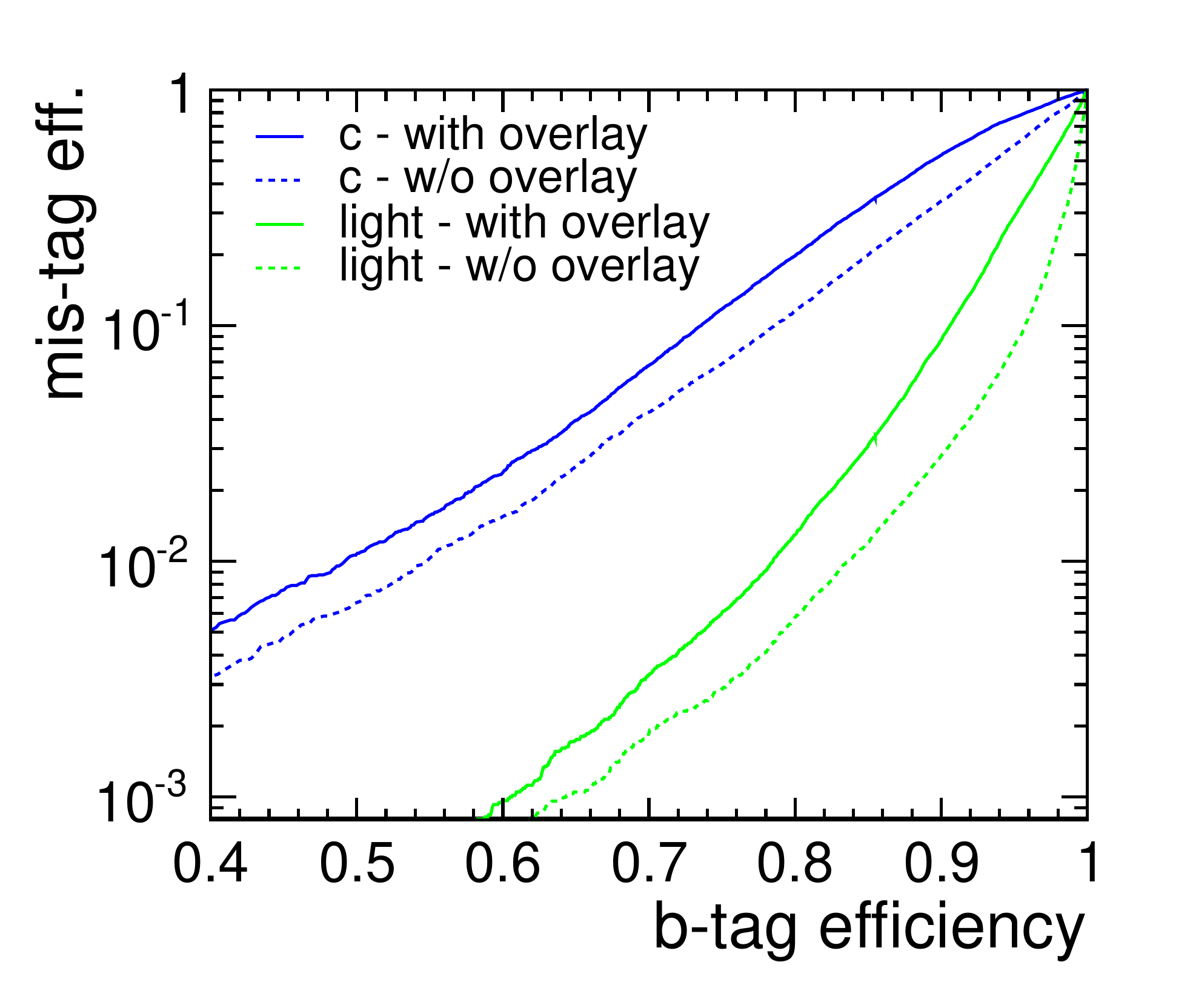}
	\includegraphics[width=.43\linewidth]{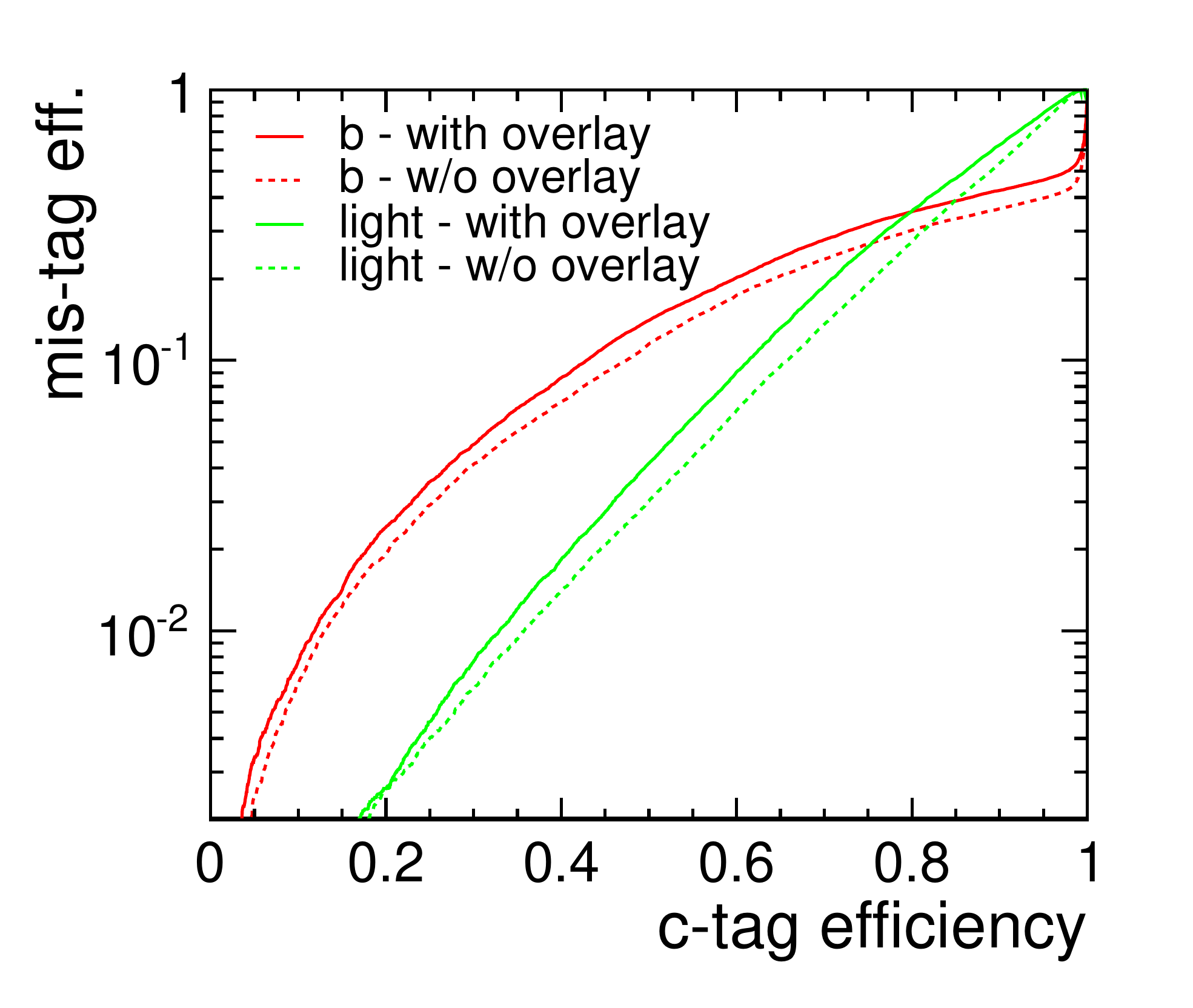}
	\caption{Purity versus efficiency curves for the LCFI flavour tagging networks in a mixed sample of $h\nu\nu$ events and $q q \nu\nu$ events. Performance for tagging $h\to b\bar{b}$ events (left). Performance for tagging $h\to c\bar{c}$ events (right).}
	\label{fig:flavourTagPerformance}
\end{figure}
\begin{wrapfigure}{r}{.4\columnwidth}
	\includegraphics[width=\linewidth]{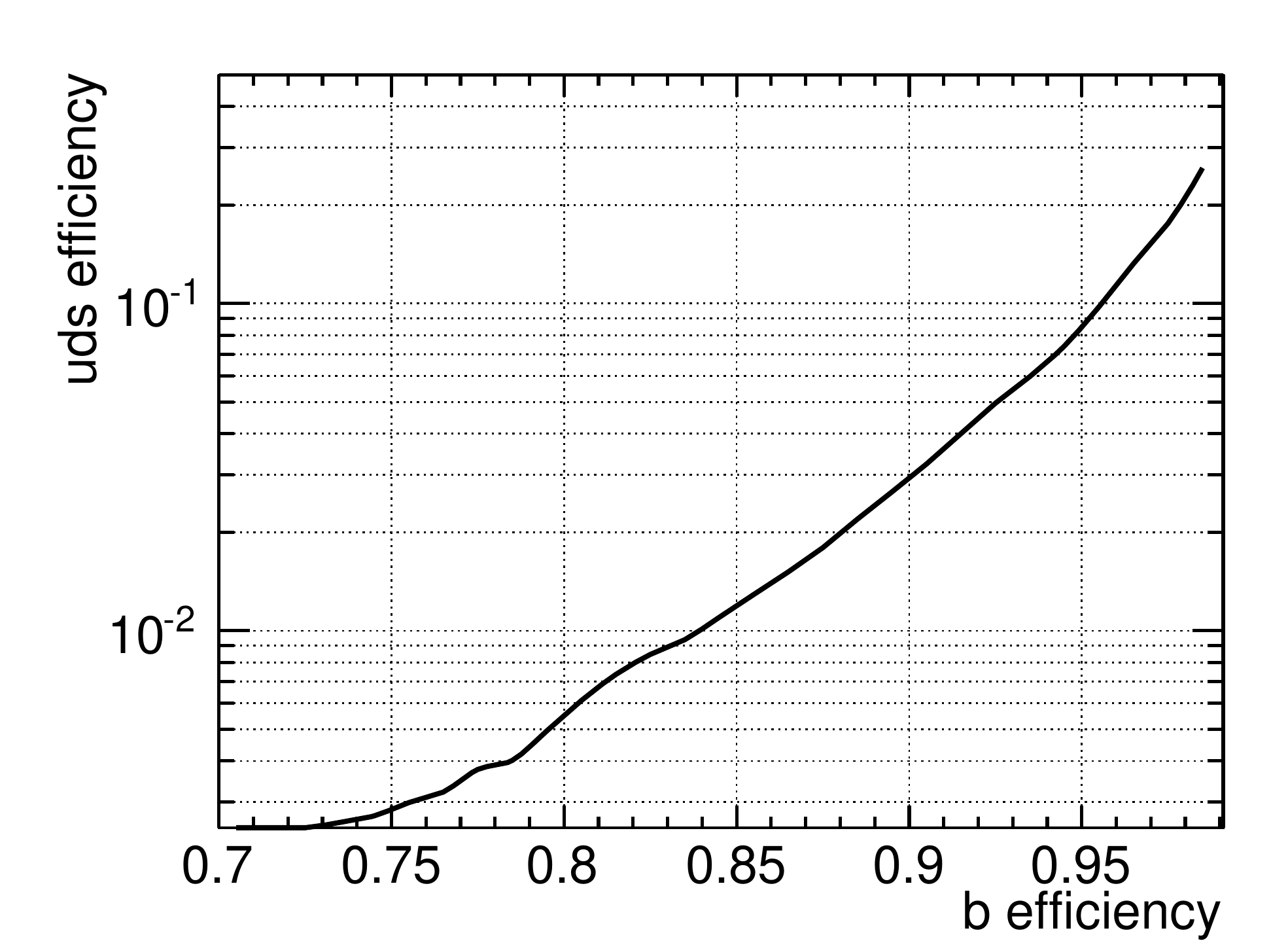}
	\caption{The performance in heavy Higgs events at a \unit[3]{TeV} CLIC}
	\label{fig:heavyHiggs}
\end{wrapfigure}
The performance of the LCFI flavour tagging package has been evaluated on the sample of light Higgs decays produced for the CLIC CDR. The flavour tagging uses three neural networks to distinguish light, c- and b-jets.
The networks are trained to distinguish b-jets from c- and light jets, c-jets from b- and light jets and c-jets from b-jets, respectively.
The nets were trained on samples with di-jets and missing energy. Figure~\ref{fig:flavourTagPerformance} shows the mis-tag efficiency for b- and c-jets of the flavour tagging network trained on b- (left) and c-jets (right) versus the tagging efficiency, where the mis-tag efficiency is the fraction of incorrectly identified jets.

Flavour tagging in events with heavy Higgs decays bears its own sets of challenges, due to the boost of the b hadrons, which results in \unit[30]{\%} of the tracks from secondary vertices originating beyond the innermost layer of the vertex detector, resulting in fewer hits. The default LCFI flavour tagging networks were augmented with track-based variables and particle identification to aid the reconstruction of $B$ decays without a reconstructed secondary vertex. They were trained on events with the same kinematic properties as the signal events. The efficiency of tagging a light jet as a b-jet versus the efficiency of correctly identifying b-jets is shown in Figure~\ref{fig:heavyHiggs}. In spite of the cited difficulties, the performance in these events is comparable to that in light Higgs events.

\subsection{Performance at \unit[500]{GeV}}
\begin{wrapfigure}{r}{.43\columnwidth}
	\includegraphics[width=\linewidth]{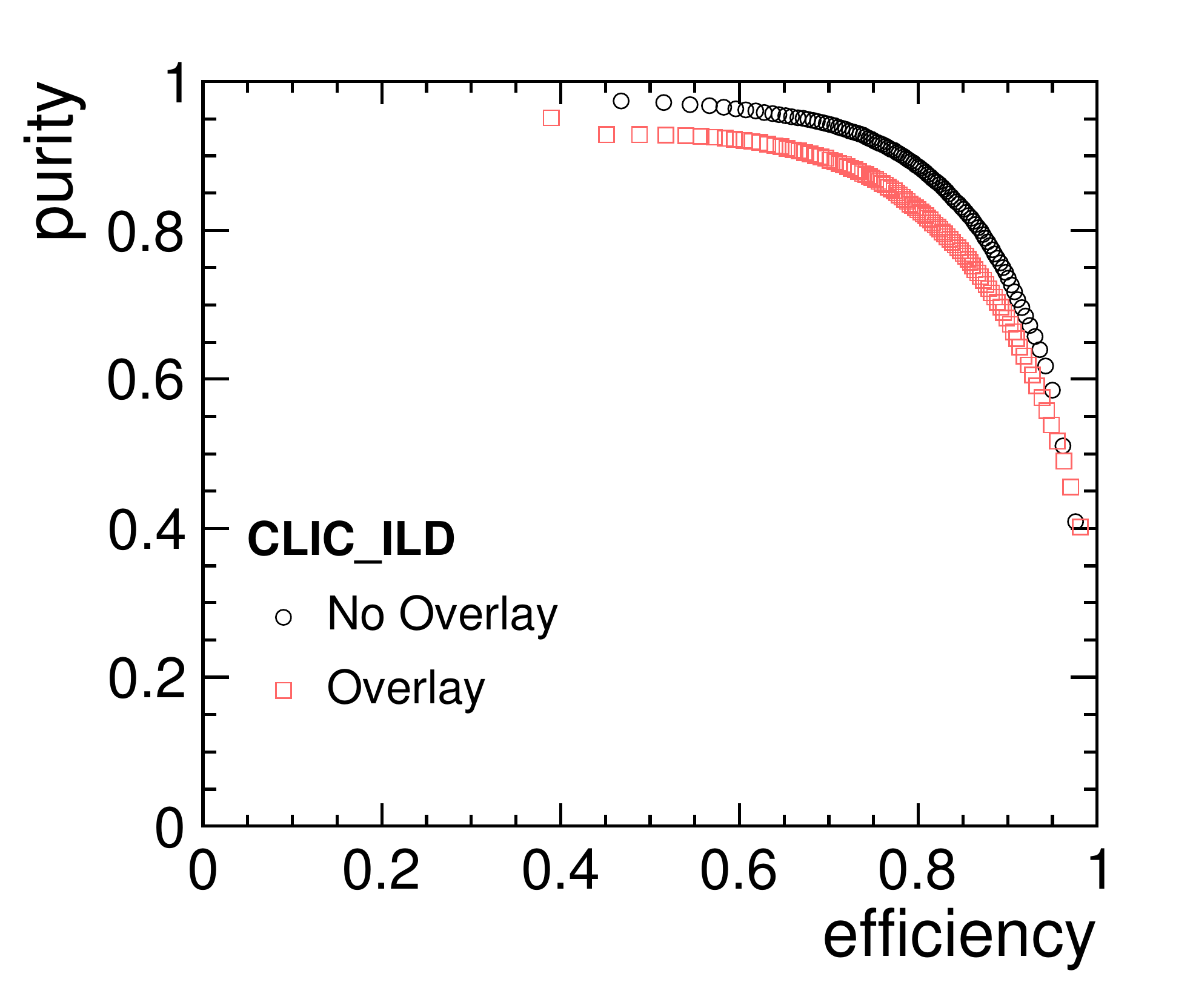}
	\caption{Flavour tagging performance in the top pair production analysis at a \unit[500]{GeV} CLIC; the black circles show the performance without background from \gghadrons events, while the red circles show the performance with this background.}
	\label{fig:flavourTagTTBar}
\end{wrapfigure}
In addition to the analysis of light Higgs decays to b and c quarks, and the analysis of heavy Higgs decays, the package was used in the study of the top pair production at \unit[500]{GeV}. The purity versus the b-tag efficiency is shown in Figure~\ref{fig:flavourTagTTBar}. The purity is the fraction of correctly identified b-jets over all tagged b-jets, evaluated in a sample of signal decays and the Standard Model background after pre-selection cuts.The analysis was carried out in the CLIC\_ILD detector concept, but due to the lower machine-induced background, the vertex detector was moved inward from its position in the detector for a \unit[3]{TeV} machine. The flavour tagging networks were trained independently for this analysis. The analysis of $t\bar{t}$ events is challenging due to the high multiplicity of the final state, where the jet finding performance impacts on the ability to find secondary vertices. Additionally, the effect of the pile-up background from \gghadrons events (Overlay) is shown for these events. The deterioration from this source is visible, but small.

\section{Summary}
We have given a brief overview over the challenges to flavour tagging at a CLIC machine and how they were addressed in the studies for the Physics and Detectors volume of the CLIC CDR. Especially the background from \gghadrons events poses challenges to the event reconstruction. Jet clustering techniques developed for hadron machines are relatively insensitive to this background. The resulting effect on the flavour tagging performance in the presented analyses is small. 

\bibliographystyle{hplain}
\bibliography{cdrbibliography}

\end{document}